\documentclass[preprint,times]{elsarticle}

\usepackage{subcaption}
\usepackage{caption}
\usepackage[T1]{fontenc}
\usepackage[utf8]{inputenc}
\usepackage{babel}
\usepackage{amsmath}
\usepackage{amssymb}

\usepackage{graphicx}
\usepackage{multirow}
\usepackage{gensymb}
\usepackage[dvipsnames]{xcolor}
\usepackage{microtype}
\usepackage{xcolor}
\usepackage[version=3]{mhchem} 
\usepackage{chemformula}[2014/04/08] 
\usepackage{booktabs}    
\usepackage{dcolumn}     
\usepackage{bm}          
\usepackage{url}
\usepackage{microtype}
\usepackage{physics}
\usepackage{units}

\usepackage[linktocpage=true,
 colorlinks=true, 
  pdfborder={0 0 0},
  linkcolor=blue,
  citecolor=red,
  filecolor=yellow,
  urlcolor=blue,
  bookmarks,
  pdfauthor={},
]{hyperref}

\usepackage[capitalise]{cleveref}

\newcommand{\alat}{A_{\text{lat}}}
\newcommand{\talat}{\tilde{A}_{\text{lat}}}
\newcommand{\nat}{{N_{\text{at}}}}
\newcommand{\tnat}{\tilde{N}_{\text{at}}}
\newcommand{\qe}{Quantum ESPRESSO}
\renewcommand{\em}{E_\text{M}}
\newcommand{\vx}{\vb{x}}
\DeclareMathOperator{\Var}{Var}

\newcommand{\pderiv}[2]{\frac{\partial #1}{\partial #2}}
\newcommand{\psderiv}[2]{\frac{\partial^2 #1}{\partial #2 ^2}}
\journal{Journal of Computational Physics: X}

\begin{document}

\begin{frontmatter}

\affiliation[0]{organization={University of Basel},
         addressline={Klingelbergstrasse 82},
         city={Basel},
         postcode={4056},
         state={Basel},
         country={Switzerland}}

\title{Efficient variable cell shape geometry optimization}

\author[0]{Moritz Gubler}
\ead{moritz.gubler@unibas.ch}
\author[0]{Marco Krummenacher}
\author[0]{Hannes Huber}
\author[0]{Stefan Goedecker}

\begin{abstract}
    A fast and reliable geometry optimization algorithm is presented that optimizes atomic positions and lattice vectors simultaneously. Using a series of benchmarks, it is shown that the method presented in this paper outperforms in most cases the standard optimization methods implemented in popular codes such as \qe{} and VASP. To motivate the variable cell shape optimization method presented in here, the eigenvalues of the lattice Hessian matrix are investigated thoroughly. It is shown that they change depending on the shape of the cell and the number of particles inside the cell. For certain cell shapes the resulting condition number of the lattice matrix can grow quadratically with respect to the number of particles. By a coordinate transformation, which can be applied to all variable cell shape optimization methods, the undesirable conditioning of the lattice Hessian matrix is eliminated.
\end{abstract}

\end{frontmatter}

\section{Introduction}\label{sec:intro}
Dynamically stable structures of periodic materials are of great scientific interest in solid state physics and materials science. Every local minimum on the potential energy surface (PES) corresponds to a dynamically stable structure. Finding local minima on the PES is a standard operation for which the availability of highly efficient algorithms is of great importance since it is used in many contexts.
For isolated clusters, a wide variety of geometry optimization methods is available. The BFGS method \cite{broyden,fletcher,goldfarb,shanno} is widely used for various optimization problems and can also be applied directly to the optimization of atomic positions. One can also use methods that were specially developed for geometry optimizations of molecules such as the modified conjugate gradient method developed by Schlegel\cite{schlegel-optimization}, the DIIS algorithm\cite{diis,csaszar_diis}, methods based on internal coordinates~\cite{schlegel_internal} or the molecular dynamics based FIRE method\cite{FIRE}. To the best of our knowledge, there are, however, no publications which discuss the optimal treatment of lattice vectors in geometry optimizations. Standard electronic structure codes contain implementations of variable cell shape geometry optimization algorithms, all with differing levels of efficiency and reliability. At first sight, the problem might seem trivial because the lattice vectors can be treated like atomic coordinates, which is what the existing implementations in electronic structure codes do. While this works well in most cases, problems arise in ill conditioned system such as molecular crystals or systems with non-compact lattice shapes. The methods presented in this paper solve these problems and therefore improve a basic tasks in the computational physics and material sciences toolbox. To improve the efficiency of variable cell shape geometry optimizations we use a change of variables which is somewhat similar to the transformation to internal coordinates\cite{schlegel_internal} that improves the efficiency of geometry optimizations for single molecules.

Periodic boundary conditions (PBC) require the addition of the degrees of freedom of the lattice to the PES:
\begin{align} \label{eq:pes_trafo}
    E(\vb x_1, \vb x_2, \dots, \vb x_\nat) \rightarrow E( \vb x_1, \vb x_2, \dots, \vb x_\nat, \alat) \nonumber\\
     = E(\alat \cdot \vb r_1, \alat \cdot \vb r_2, \dots, \alat \cdot \vb r_\nat, \alat)
\end{align}
\cref{eq:pes_trafo} shows this modification where $\vb x_i$ are the Cartesian coordinates of an atom, $\alat = (\vb{a} | \vb{b} | \vb{c})$ is the three by three matrix containing the three lattice vectors $\vb{a},\, \vb{b},\, \vb{c}$ and $r_i$ are the reduced (fractional) coordinates of the atoms that specify the atomic positions in terms of the lattice vectors. All components of the reduced coordinates $r_i$ are in the interval $[0, 1]$. The form of the PES for PBC in \cref{eq:pes_trafo} introduces some complications for finding local minima on the PES. The coordinates $\tilde{x}_i$ and $\tilde{A}$ of a local minimum on the PES with PBC are defined by \cref{eq:locmin} and a curvature condition.
\begin{align} \label{eq:locmin}
    \frac{\partial E (\vb x_1, \dots, \vb x_\nat, \alat)}{\partial \alat} \Bigr|_{\substack{\vb x_i=\tilde{\vb x}_i\\\alat=\talat}} &= 0 \nonumber \\
    \frac{\partial E (\vb x_1, \dots, \vb x_\nat, \alat)}{\partial \vb x_i}  \Bigr|_{\substack{\vb x_i=\tilde{\vb x}_i\\\alat=\talat}} &= 0 \quad \forall\, i = 1, \dots, \nat
\end{align}
In order to find a local minimum on the PES it is not sufficient to adjust the atomic positions $\vb x_i$. It is necessary to relax the lattice matrix $\alat$ as well. 
The implicit dependence of atomic positions on the lattice vectors in \cref{eq:pes_trafo} complicates the extension of geometry relaxation algorithms from free to periodic systems. This article extends the stabilized quasi Newton method (SQNM) developed for free structures by Schaefer et al.\cite{schaefer_geopt} to periodic structures and proposes a new step size control method. The resulting method will then be tested against several implementations of periodic geometry relaxation algorithms included in standard electronic structure codes.

The largest eigenvalue of the Hessian matrix ($\lambda_{\text{max}} $) contains important information for geometry optimizations. Its inverse is the optimal step size for steepest descent algorithms. For the Hessian matrix of the atomic positions, $\lambda_{\text{max}}$ is not dependent on the system size. The smallest eigenvalue of the Hessian matrix of the atomic position decreases as the system is increased in size\cite{free_step_size}.
In this paper we will present an analytical calculation of the largest eigenvalue of the lattice part of the Hessian matrix which demonstrates that $\lambda_{\text{max}}$ is indeed dependent on the number of particles. In order to corroborate this result $\lambda_{\text{max}}$ is calculated numerically using energy and forces obtained with a force-field\cite{silicon-ff} and \qe{}\cite{qe-2009,qe-2017,qe-2020}.

The results of the analytical calculation, combined with a thorough numerical investigation of the lattice Hessian matrix, show that there are some serious ill-conditioning problems in variable cell shape geometry optimizations. In \cref{sec:optalg} we present a new variable cell shape geometry optimization method that is not affected by these problems.

\section{Dependence of curvature of the PES with respect to the lattice parameters on the number of particles and cell shapes}\label{chap:lat_dep}
In this section we present both analytical and numerical calculations of eigenvalues of the Hessian matrix of the lattice part of the PES.

\subsection{Analytic calculation} \label{sec:ana-calc}
Here, three scenarios are presented where the eigenvalues of the lattice part of the Hessian matrix can be calculated analytically. 

In the first case, a cubic cell is considered with cell length $a$. Here, $a$ is varied while the particle density remains the same. In this scenario, the eigenvalue of the lattice part of the Hessian matrix is the second derivative of the PES with respect to the cell parameter $a$. Since the total energy is an extensive quantity, the energy of two crystals with cell parameters $a$, $\nat$ and $\tilde{a} = m \cdot a$, $\tnat = m^3 \nat$ is related by $E(a, \nat) = \frac{a^3}{\tilde{a}^3} E(\tilde{a}, \tnat) =\frac{1}{m^3} E(\tilde{a}, \tnat)$. Differentiating both sides twice with respect to $a$ leads to $\frac{\partial^2 E(a, \nat)}{\partial a^2} = \frac{1}{m} \frac{\partial^2 E(\tilde{a}, \tnat)}{\partial \tilde{a}^2}$. Therefore, $\frac{\partial^2 E(a)}{\partial a^2}$ is proportional to the lattice parameter $a$. Because the particle density remains constant, $\frac{\partial^2 E(a)}{\partial a^2}$ is also proportional to $\nat^{\frac{1}{3}}$.

In the second scenario, we consider an orthorombic cell where the cell lengths are $a$, $a$ and $b$. Once again the energies of two crystals with cell parameters $(a, a, b)$ and $(\tilde{a} = m \cdot a, \tilde{a}=m \cdot a, b)$ where the second crystal cell is scaled by a factor $m$. The number of particles is chosen such that the density remains constant and therefore, $\tnat = m^2 \nat$. The energies are related by $ E(a, b, \nat) = \frac{a^2 b}{\tilde{a}^2 b} E(\tilde{a}, b, \tnat) = \frac{1}{m^2} E(m \cdot a, b, m^2 \nat)$. The second derivatives are $ \frac{\partial^2 E(a, b, \nat)}{\partial a^2} = \frac{\partial^2 E(\tilde{a}, b, \tnat)}{\partial \tilde{a}^2}$, $ \frac{\partial^2 E(a, b, \nat)}{\partial b^2} =\frac{1}{m^2} \frac{\partial^2 E(\tilde{a}, b, \tnat)}{\partial b^2} $ and $\frac{\partial^2 E(a, b, \nat)}{\partial a \partial b} =\frac{1}{m} \frac{\partial^2 E(\tilde{a}, b, \tnat)}{\partial \tilde{a} \partial b}$. Since the number of atoms is proportional to $m^2$ one can write the Hessian matrix $H_{\text{lat}}$ in terms of the number of atoms: 
\begin{equation} \label{eq:h_lat_flat}
H_{\text{lat}} = 
\begin{pmatrix}
    \frac{\partial^2 E(a, b)}{\partial a^2} = const. & \frac{\partial^2 E(a, b)}{\partial a \partial b} \sim \sqrt{\nat}\\
    \frac{\partial^2 E(a, b)}{\partial a \partial b} \sim \sqrt{\nat} & \frac{\partial^2 E(a, b)}{\partial b^2} \sim \nat
\end{pmatrix}.
\end{equation}
The first eigenvalue of \cref{eq:h_lat_flat} grows asymptotically linear with respect to the number of atoms and the second eigenvalue is asymptotically constant. This can be seen by plugging the Hessian matrix from \cref{eq:h_lat_flat} into the formula that calculates eigenvalues of two by two matrices.

Lastly, we consider two orthorombic cells with cell lengths $(a, b, b)$ and $(\tilde{a}, b, b)=(m\cdot a, b,b)$. Once again the number of particles is scaled such that the particle density remains constant: $\tnat = m \nat$. Their energies are given by $E(a, b, \nat) = \frac{1}{m} E(m \cdot a,b, \tnat)$. The calculation of the Hessian matrix is analogous to the derivation of \cref{eq:h_lat_flat} and the resulting Hessian matrix has the form
\begin{equation} \label{eq:h_lat_long}
H_{\text{lat}} = 
\begin{pmatrix}
    \frac{\partial^2 E(a, b)}{\partial a^2} \sim \frac{1}{\nat} & \frac{\partial^2 E(a, b)}{\partial a \partial b} =const.\\
    \frac{\partial^2 E(a, b)}{\partial a \partial b} = const. & \frac{\partial^2 E(a, b)}{\partial b^2} \sim \nat
\end{pmatrix}.
\end{equation}
In this case, one eigenvalue of \cref{eq:h_lat_long} grows asymptotically linear and the other one approaches $\frac{1}{\nat}$ asymptotically.

\begin{figure}
\centering
    \includegraphics[width=8.6cm]{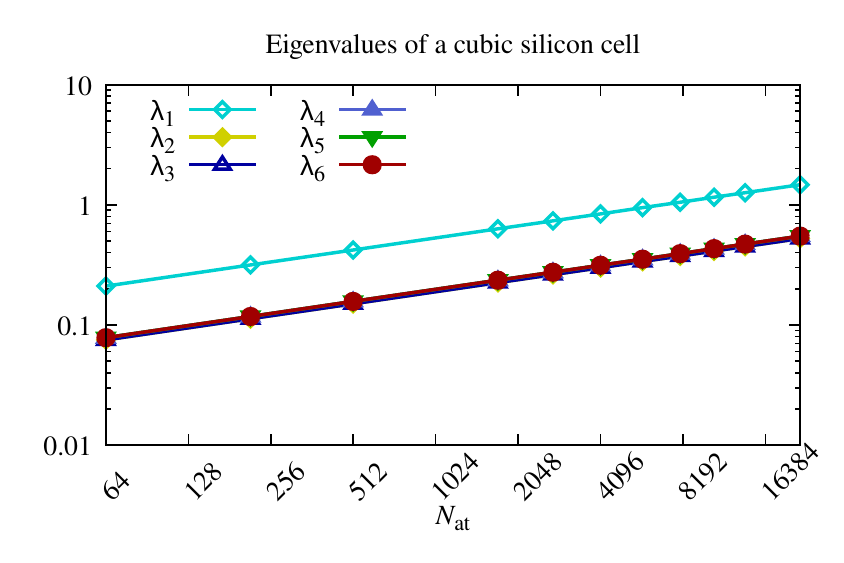}
    \caption{A log plot of all six eigenvalues of the Hessian matrix of a silicon force field~\cite{silicon-ff}. $\lambda_{2-6}$ are degenerate and not all lines are visible. A fit gives a slope of nearly exactly $\frac{1}{3}$ for all eigenvalues.}
    \label{fig:hess-cube}
\end{figure}

\subsection{Numerical evaluation of the Hessian matrix and its eigenvalues} \label{sec:numeval}
In this section we confirm our analytical results by numerical experiments. The eigenvalues of the lattice part of the Hessian matrix are calculated for increasing numbers of particles for the cell shapes, studied in the previous section. In contrast to the analytical calculation, all six degrees of freedom of the lattice Hessian matrix are now incorporated in the numerical test of the model developed in \cref{sec:ana-calc}.

The cubic growth was done using a force field developed by Bazant et al.~\cite{silicon-ff}. The ground state that was represented in a cell containing 64 atoms was then expanded equally in all directions and the lattice part of the Hessian matrix was calculated numerically.
\cref{fig:hess-cube} shows the results of this calculation. The fitted lines in \cref{fig:hess-cube} all have a slope of nearly exactly $\frac{1}{3}$ which shows that all eigenvalues are proportional to $\nat^{\frac{1}{3}}$.

\begin{figure}[t]
\centering
    \includegraphics[width=8.6cm]{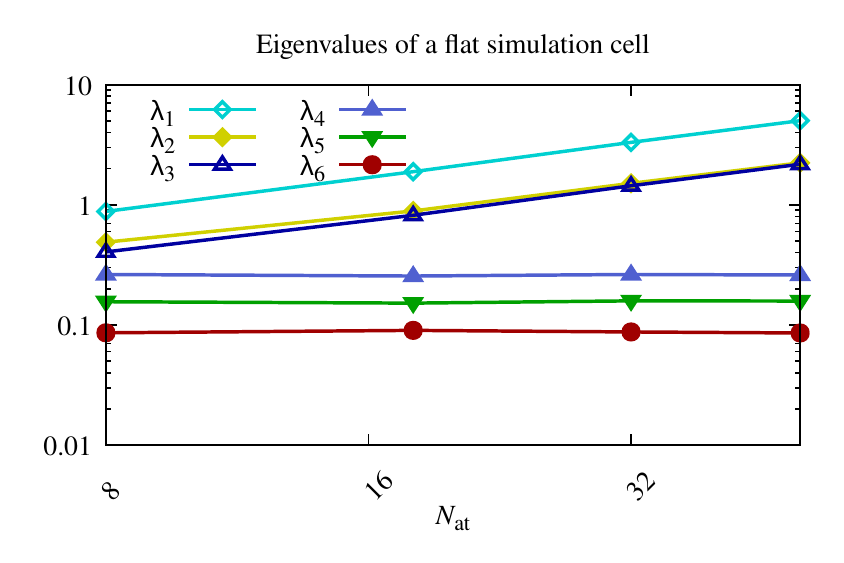}
    \caption{Log plot of the eigenvalues of a cell expansion in two directions. The fitted slopes of $\lambda_{1-6}$ are 0.95, 0.87, 0.94, 0.00, 0.01 and 0.00 respectively.
    \label{fig:hess-flat}}
\end{figure}

\begin{figure}[t]
    \centering
    \includegraphics[width=8.6cm]{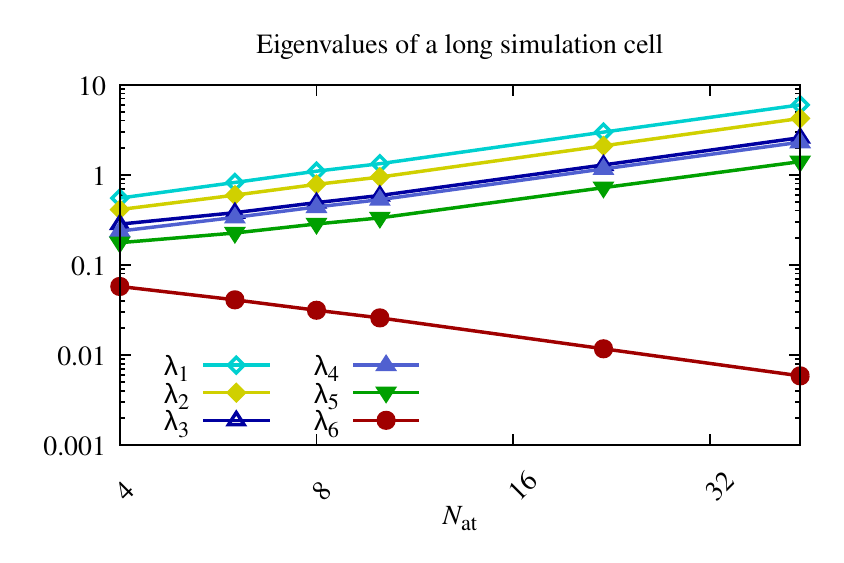}
    \caption{Log plot of the eigenvalues of a cell expansion in a single direction. A cell of this kind is pictured in \cref{fig:si-long}. The fitted slopes of $\lambda_{1-6}$ are 1.00, 1.00, 0.98, 0.98, 0.93 and -0.91 respectively.}
    \label{fig:hess-long}
\end{figure}

The other two cell shapes are analyzed by expanding a diamond cell in one and two directions which resulted in long and flat cells. The energies are calculated on the DFT level using the \qe{} package and the Hessian matrix was calculated numerically using finite differences. \cref{fig:hess-flat,fig:hess-long} show the results of this calculation. In the flat simulation cell three eigenvalues depend linearly on the number of particles and the remaining three are constant. For the long cell where the lattice cell was expanded in only one direction five eigenvalues grow linear with respect to the number of particles and one is proportional to $\frac{1}{\nat}$. The analytic model which is derived in \cref{sec:ana-calc} does not perfectly predict the numerically calculated eigenvalues shown in \cref{fig:hess-flat,fig:hess-long}. Since the model from \cref{sec:ana-calc} only predicts the asymptotic behaviour of the eigenvalues one would have to consider very large cells for results that tend to the asymptotic value. Another challenge is the noise which is inherently present in all DFT calculations which makes it challenging to accurately calculate derivatives using finite differences. We repeated the calculations for the eigenvalues from \cref{fig:hess-flat,fig:hess-long} using a force field and more atoms which results in a nearly perfect agreement between the model presented in \cref{sec:ana-calc} and the numerically calculated eigenvalues.



\subsection{Conditioning of the lattice Hessian matrix}
The condition number of an optimization problem is defined by the ratio between the largest and the smallest eigenvalue of the Hessian matrix. In \cref{sec:ana-calc,sec:numeval} the lattice cells are expanded equally in all directions such that the proportions of the cells remained the same. The resulting condition number is constant with respect to the number of particles. This can be seen from \cref{fig:hess-cube} where all the eigenvalues are proportional to $\nat ^{\frac{1}{3}}$. Hence the ratio is constant.

When the lattice cell is not expanded equally in all directions, the condition number does not remain constant with respect to the number of particles. The condition number that grows linearly with respect to the number of particles in a flat cell and in long cells the condition number can even grow quadratically with respect to the number of particles in the simulation cell.

\section{Lattice updates}\label{sec:latup}
The implicit dependence of the atomic positions on the lattice vectors introduces some difficulty to update lattice vectors. In \cref{sec:intro} it is shown that the Cartesian positions of all atoms change when the lattice vectors are changed for fixed reduced coordinates. This behaviour is problematic when standard quasi Newton methods based on Cartesian coordinates are used to optimize the energy on the PES. Quasi Newton methods extract curvature information based on the differences between input vectors and gradients from previous iterations. When a lattice vector update changes the atomic positions, the quasi Newton algorithm is disturbed and extracts poor curvature information. This has to be avoided. In this work we present several possibilities to overcome this problem.

\subsection{Lattice update in reduced atomic positions} \label{sec:redlatup}
Instead of optimizing the Cartesian $\vb x_i$ coordinates of the PES one can also optimize the reduced coordinates $\vb r_i$. 
\begin{align}
    \tilde{E} (\vb r_1, \vb r_2, \dots, \vb r_\nat, A)  \nonumber\\
    &= E(A \cdot \vb r_1, A \cdot \vb r_2, \dots, A \cdot \vb r_\nat, A) \nonumber\\
     &= E (\vb x_1, \vb x_2, \dots, \vb x_\nat, A) \\ 
    \frac{\partial \tilde{E}}{\partial \vb r_i} &=\underbrace{\frac{\partial \vb x_i}{\partial \vb r_i}}_{\alat} \frac{\partial E}{\vb x_i} = \alat  \cdot\frac{\partial E}{\vb x_i} 
\end{align}
Since $\frac{\partial \tilde{E}}{\partial \vb r_i} = 0 \iff \frac{\partial E}{\partial \vb x_i} = 0$ and since $\alat$ can be chosen to be positive definite, each local minimum of $E$ corresponds to a local minimum of $\tilde{E}$.

This update mechanism is also problematic. Usually, the cell vectors differ in length and these differences can even be arbitrarily large. 
The transformation of the Cartesian coordinates to the reduced coordinates maps all entries of the atomic coordinates into the interval $[0,1]$. If the lattice vector is long the distance between the mapped point is small which leads to an increase of the curvature. Hence the condition number grows for cells whose lattice vectors have different lengths.
\subsection{Quasi Cartesian coordinates}
By introducing quasi Cartesian coordinates it is possible to 
eliminate the increase of the condition number for flat or elongated cells.
Let $A_0$ be the lattice matrix from the initial non-relaxed cell and $\vb q_i = A_0 \cdot \alat^{-1} \cdot \vb x_i = A_0 \cdot \vb r_i$ the quasi Cartesian coordinates describing the position of atom $i$. Instead of optimizing the usual PES which depends on the Cartesian coordinates $\vb x_i$ and the lattice matrix $\alat$ we now optimize 
    \begin{align}
        \tilde{E} ( & \vb q_1, \dots, \vb q_\nat, \alat) \nonumber \\ 
        &= E(\alat \cdot A_0^{-1} \cdot \vb q_1, \dots, \alat \cdot A_0^{-1} \cdot \vb q_\nat, \alat) \nonumber \\
        &= E (\vb x_1, \dots, \vb x_\nat, \alat).
    \end{align}
    The derivatives of $\tilde{E}$ can be calculated using the chain rule:
    \begin{equation}
        \frac{\partial \tilde{E}}{\partial \vb q_i} = \frac{\partial \vb x_i}{\partial \vb q_i} \frac{\partial E}{\partial \vb x_i} = \alat \cdot A_0^{-1} \frac{\partial E}{\partial \vb x_i}.
    \end{equation}

\section{Improving the condition number of the lattice optimization}
\subsection{Supercell lattice derivatives}
In \cref{chap:lat_dep} it was shown, that the condition number is very large when very long or very flat cells are considered. This problem can be overcome by expanding the non-compact cell into a supercell where all of the cell vectors have the same length and calculating the lattice derivatives of this supercell. It is not necessary to calculate the lattice derivatives directly for this supercell. Using the fact that the stress tensor $\sigma$ remains invariant under this expansion and the relation 
\begin{equation}
    \frac{\partial E}{\partial \alat} = -\det (\alat) \cdot \sigma \cdot (\alat^{-1})^\mathrm{T}
\end{equation}
it is possible to transform the lattice derivatives of the small, original cell to the bigger hypothetical cell. A good choice for the hypothetical cell $\tilde{A}$ is 
\begin{equation}\label{eq:super-lat}
    \tilde{A} \sim \alat \cdot     \begin{pmatrix}
            \frac{1}{\norm{\vb{a}_0}} & 0 & 0\\
        0 & \frac{1}{\norm{\vb{b}_0}} & 0\\
        0 & 0 & \frac{1}{\norm{\vb{c}_0}}
    \end{pmatrix}
\end{equation}
where $\vb{a}_0$, $\vb{b}_0$ and $\vb{c}_0$ are the lengths of the initial lattice vectors. \cref{eq:super-lat} transforms $\alat$ to $\tilde{A}$ where all lattice vectors have approximately the same length.
The lattice derivatives $\tilde{A}$ are proportional to 
\begin{equation}\label{eq:dsuper-lat}
\pderiv{E}{\alat} \cdot \begin{pmatrix}
            \frac{1}{\norm{\vb{a}_0}} & 0 & 0\\
        0 & \frac{1}{\norm{\vb{b}_0}} & 0\\
        0 & 0 & \frac{1}{\norm{\vb{c}_0}}
    \end{pmatrix}.
\end{equation}
The size of $\tilde{A}$ is somewhat arbitrary. Because of that only the proportionality of \cref{eq:super-lat,eq:dsuper-lat} is given. A good choice for the size of supercells is presented in \cref{sec:lattice_trafo_cetors}

\subsection{Resulting transformation of lattice vectors} \label{sec:lattice_trafo_cetors}
In order to improve the condition number of the lattice matrix the lattice vectors should be transformed such that the resulting lattice derivatives have the form of \cref{eq:dsuper-lat}. The transformation
\begin{equation}
    \tilde{A} = \gamma \cdot \, \alat \cdot     \begin{pmatrix}
            \frac{1}{\norm{\vb{a}_0}} & 0 & 0\\
        0 & \frac{1}{\norm{\vb{b}_0}} & 0\\
        0 & 0 & \frac{1}{\norm{\vb{c}_0}}
    \end{pmatrix}
\end{equation}
results in a lattice derivative of the desired form where $\gamma$ is a scaling factor that will be set appropriately later in this section. Now, instead of optimizing $E(\vb x_1, \vb x_2, \dots, \vb x_\nat, \alat)$, we optimize
\begin{align}
    &\tilde{E} (\vb q_1, \vb q_2, \dots, \vb q_\nat, \tilde{A}) \nonumber \\
    &= E(\vb x_1, \vb x_2, \dots, \vb x_\nat, \tilde{A} \cdot \frac{1}{\gamma} \cdot
    \begin{pmatrix}
        \norm{\vb{a}_0} & 0 & 0\\
        0 & \norm{\vb{b}_0} & 0\\
        0 & 0 & \norm{\vb{c}_0}
    \end{pmatrix}) \label{eq:pes_lat_trafo}
\end{align}
with respect to $\tilde{A}$. The lattice derivative of \cref{eq:pes_lat_trafo} is given by:
\begin{equation}
    \frac{\partial \tilde{E}}{\partial \tilde{A}} = \frac{\partial \alat}{\partial \tilde{A}} \frac{\partial E}{\partial \alat} = \frac{\partial E}{\partial \alat} \cdot \frac{1}{\gamma} \cdot
    \begin{pmatrix}
        \norm{\vb{a}_0} & 0 & 0\\
        0 & \norm{\vb{b}_0} & 0\\
        0 & 0 & \norm{\vb{c}_0}
    \end{pmatrix}.
\end{equation}
Since the lattice derivatives from the modified potential energy surface of \cref{eq:pes_lat_trafo} are now the derivatives of a supercell were the lengths of the lattice vectors are equal, the eigenvalues of the lattice Hessian matrix of \cref{eq:pes_lat_trafo} now behave like the ones shown in \cref{fig:hess-cube} where all of the eigenvectors are proportional to $\nat^{\frac{1}{3}}$. 
Therefore, the condition number of the lattice part of the Hessian matrix is constant with respect to the system size. Since the smallest eigenvalue of the position part of the Hessian matrix decreases with increasing system size, the overall condition number of the entire Hessian matrix is worsened by the $\nat^{\frac{1}{3}}$ dependence of the eigenvalues from the lattice part of the Hessian matrix. Ideally, the eigenvalues of the lattice part of the Hessian matrix should also be constant. In this way they will not increase the condition number of the entire Hessian matrix. This can be achieved by choosing $\gamma$ properly. When $\gamma \sim \sqrt{\nat}$, the lattice eigenvalues do not depend on the number of particles anymore. This is derived in the appendix A.

\begin{figure}
    \centering
    \includegraphics{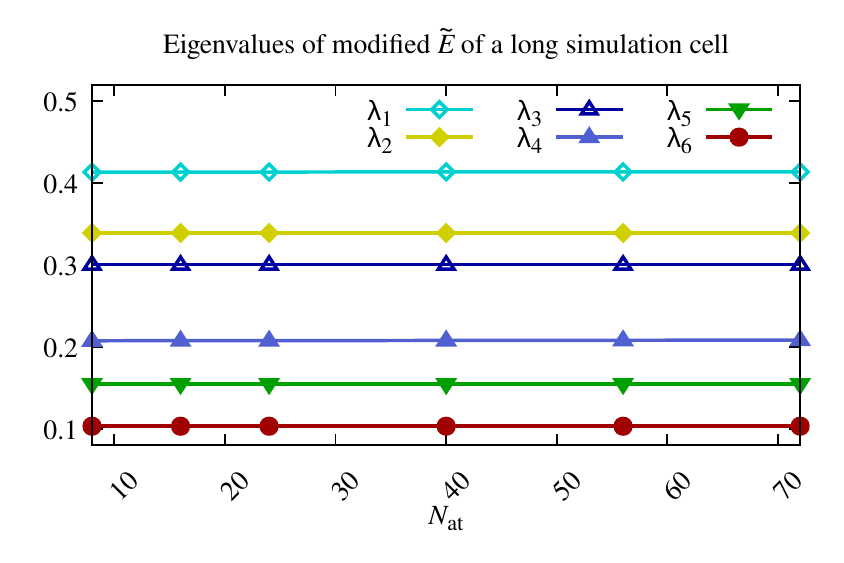}
    \caption{The modified eigenvalues of the lattice Hessian matrix for $w=2$ for a long silicon cell. The eigenvalues and the condition number are now invariant with respect to the system size. The problematic spectrum displayed in \cref{fig:hess-long} is corrected.}
    \label{fig:longcell-precon}
\end{figure}

Finally, one has to choose a parameter $w$ to set the magnitude of $\gamma= w \cdot \sqrt{\nat}$. $w$ determines the overall curvature of the lattice Hessian matrix and should be chosen such that the lattice and position curvature are similar. Based on our experience $w$ should be between 1 and 2 when atomic units are used. \cref{fig:longcell-precon} shows that the eigenvalues of the Hessian matrix of the transformed lattice vectors are indeed constant with respect to the system size.

\section{Summary of the transformations that eliminate ill-conditioning} \label{sec:optalg}

Our theoretical analysis of the periodic potential energy surface shows the the condition number is improved when the lattice and position variables are transformed. Therefore, one should optimize $\tilde{E}(\vb q_1, \dots, \tilde{A})$ with respect to $\vb q_i$ and $\tilde{A}$. Each minimum on $\tilde{E}$ corresponds to a minimum on $E$.
The two transformations are: $\vb q_i = A_0 \cdot \alat^{-1} \cdot \vb x_i = A_0 \cdot \vb r_i$ and 
\begin{equation}
    \tilde{A} = w \cdot \sqrt{\nat} \cdot \, \alat \cdot      \begin{pmatrix}
            \frac{1}{\norm{\vb{a}_0}} & 0 & 0\\
        0 & \frac{1}{\norm{\vb{b}_0}} & 0\\
        0 & 0 & \frac{1}{\norm{\vb{c}_0}}
    \end{pmatrix}
\end{equation}
with the derivatives being $\frac{\partial \tilde{E}}{\partial \vb q_i} = \alat \cdot A_0^{-1} \frac{\partial E}{\partial \vb x_i}$ and 
\begin{equation}
        \frac{\partial \tilde{E}}{\partial \tilde{A}} = \frac{\partial E}{\partial \alat} \cdot \frac{1}{w \cdot \sqrt{\nat}} \cdot
    \begin{pmatrix}
        \norm{\vb{a}_0} & 0 & 0\\
        0 & \norm{\vb{b}_0} & 0\\
        0 & 0 & \norm{\vb{c}_0}
    \end{pmatrix}.
\end{equation}
with $w$ being a value between one and two atomic units. After these transformations of the potential energy surface any minimization method can be applied to $\tilde{E}$. The quasi newton method only sees the the variables $\vb q_i$ and $\tilde{A}$. After a stationary point is found i.e. $\pderiv{\tilde{E}}{\vb q_i}=0$ and $\pderiv{\tilde{E}}{\tilde{A}}=0$, the coordinates are transformed back and $\vb x_i$ and $\tilde{A}$ are returned.

We use the SQNM method~\cite{schaefer_geopt} as our optimizer, since it is reliable, efficient and developed especially for potential energy surfaces. Any other gradient based optimization method can also be accelerated by the transformations described in this paper.

\section{Additional useful functionalities of our geometry optimizer}
The equations and functionalities derived in this chapter can be useful even when other optimization methods are used and they may be applied to fixed cell optimization problems. Whenever a vector $\vx$ is referred to, we assume that the quasi cartesian coordinates $q_i$ and the transformed lattice vectors $\tilde{A}$ are concatenated into this vector.

\subsection{Improved step size control for the SQNM method}
The step size control mechanism is improved in our implementation of the variable cell SQNM (vc-SQNM) method. Instead of using the angle between the original gradient and the preconditioned gradient as a feedback as described in the original SQNM\cite{schaefer_geopt} method, we use the gain ratio, a quantity which is used in trust region methods\cite{num-opt} to adjust the trust radius. It is defined by:
\begin{equation}
    \rho = \frac{\text{actual gain}}{\text{expected gain}} = \frac{E_k - E_{k+1}}{m_k - m_{k+1}}
\end{equation}
where $E_k$ is the energy at iteration $k$ and $m_k$ is the approximated energy that can be calculated using the estimated Hessian matrix in quasi Newton methods. The approximated Hessian matrix outside of the significant subspace in the SQNM method is diagonal and all diagonal elements are equal to the inverse stepsize of the current iteration. A good way to adapt the step size is to increase the step size if the gain ratio is bigger than one and to decrease it if the gain ratio is smaller than 0.5.

\subsection{Estimation of a lower bound for the ground state energy}
\begin{figure}
    \centering
    \includegraphics{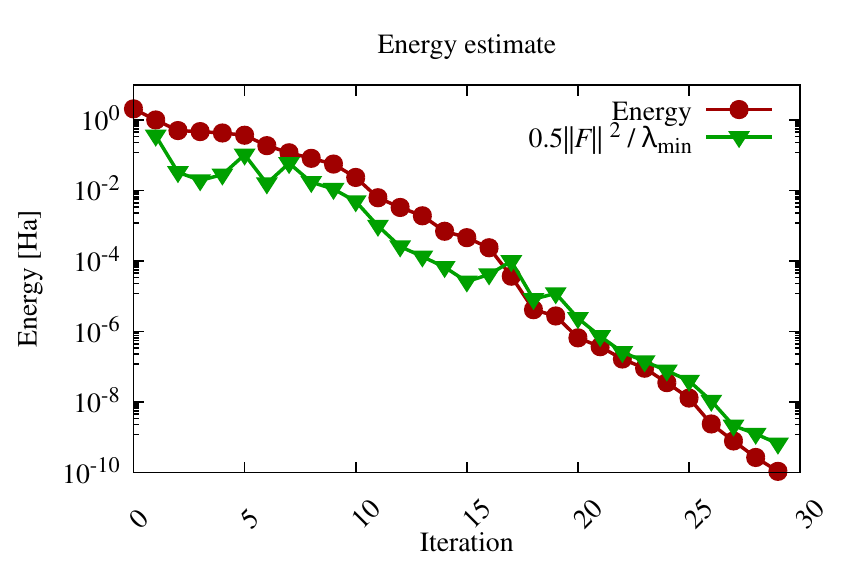}
    \caption{Exact energy difference to ground state and the estimate of the same quantity, $\frac{|| F(\vb x) ||^2}{2\lambda_\text{min}}$, plotted against the number of geometry optimization iterations.}
    \label{fig:ener_lower_lim}
\end{figure}

A feature of quasi Newton methods is their estimate of the Hessian matrix of the PES which is systematically improved during the geometry optimization. This can be utilized to estimate a lower bound for the energy of the local minimum that the geometry optimization is approaching. Let $\em$ be the quadratic model of the PES that the quasi Newton method generates. It has the form
\begin{equation}\label{eq:model_pes}
    \em (\vb x) = E_0 + (\vb x - \vb x_0)^\mathrm{T} \underbrace{\nabla E(\vb x_0)}_{=0} + \frac{1}{2} (\vb x - \vb x_0)^\mathrm{T} A (\vb x - \vb x_0).
\end{equation}
$\vb x_0$ is the vector holding the coordinates of the local minimum, $A$ is the current quasi Newton estimate of the symmetric Hessian matrix and $E_0$ is the energy of the local minimum. $\nabla E(\vb x_0) = 0$ because there are no residual forces in the local minimum. The gradient of $\em$ can be written as $\nabla \em (\vb x) = A (\vb x - \vb x_0)$. $\em$ can be expressed in terms of its own gradient: $\em (\vb x) = E_0 + \nabla \em (\vb x) A^{-1} \nabla \em (\vb x)$. When the quadratic model $\em$ captures the shape of the PES well, $\em(\vb x) \approx E(\vb x)$ and $\nabla \em(\vb x) \approx  - \vb F(\vb x)$. Therefore, 
\begin{equation}
    E(\vb x) \approx E_0 + \frac{1}{2} \underbrace{\vb F(\vb x)^\mathrm{T} A^{-1} \vb F(\vb x)}_{\leq \frac{|| \vb F(\vb x) ||^2}{\lambda_\text{min}}}
\end{equation}
where $\lambda_\text{min}$ is the smallest eigenvalue of $A$. This leads to the estimate of the lower bound of the ground state energy 
\begin{equation}\label{eq:ener_diff}
    E_0 \geq E(\vb x) - \frac{|| \vb F(\vb x) ||^2}{2 \lambda_\text{min}}. 
\end{equation}
\cref{fig:ener_lower_lim} shows the energy difference to the ground state and the estimated energy difference that was calculated using \cref{eq:ener_diff} versus the number of geometry optimization iterations. 

\subsection{Detection of noise level in the forces} \label{sec:noise-force}
In most atomistic simulations, the PES is calculated by some kind of iterative DFT procedure  or by stochastic Monte Carlo methods. In both cases solutions are not exact - they are always approximate and contain therefore some numerical noise. Typically, the error in the forces e.g. the gradients of the PES is larger than the error that is present in the potential energy. A large error in the forces of the PES is undesirable in the context of geometry optimizations and can worsen the performance of the geometry optimization dramatically. Here we present a measure to detect and quantify noise.

Noisy forces can be modelled in the following way:
\begin{equation}
    F_i = \tilde{F}_i + \epsilon_i  .
\end{equation}
$F_i$ is the force component obtained using the numerical calculation, $\tilde{F}_i$ is the true value of the force component and $\varepsilon_i$ is a random number drawn from the probability distribution $\mathcal{E}_i$ with zero mean and an unknown variance $\Var(\mathcal{E}_i)$. Note that $\mathcal{E}_i$ may be a different, independent distribution for each force component. The true force components $\tilde{F}_i$ must sum up to zero. Therefore, $S = \sum F_i = \sum \varepsilon_i$. Using Bienaymé's identity~\cite[p. 116]{klenke2007probability} (which states that $\Var(\sum \mathcal{E}_i) = \sum \Var(\mathcal{E}_i)$ in this case), the variance of $S$ can be expressed as $\Var(S) = \sum \Var(\mathcal{E}_i)$. The variance of $S$ can now be estimated using
\begin{equation}
    \Var(S) = \frac{1}{3} \sum_{j=1}^3 \left(\sum_{i=1}^\nat F_{i, j}\right)^2 \geq \Var(\mathcal{E}_l) \quad \forall l=1, \dots, \nat.
\end{equation}
So far we have found a rather weak upper bound of the variances of $\mathcal{E}_i$. If one also assumes that all $\Var (\mathcal{E}_i)$ are equal then
$
    \Var(\mathcal{E}_i) = \frac{\Var(S)}{\nat} \quad \forall i = 1, \dots, \nat.
$
The standard deviation $\sigma$ of $\mathcal{E}_i$ is a measure of the noise level present in the forces. $\sigma$ can be calculated using the following relation:
\begin{equation}
    \sigma = \sqrt{\frac{\Var(S)}{\nat}} = \sqrt{ \frac{1}{3\nat} \sum_{j=1}^3 \left(\sum_{i=1}^\nat F_{i, j}\right)^2 }.
\end{equation}
When $\sigma$ is comparable to the magnitudes of the force components, the convergence of geometry optimizations will come to a standstill because the forces are dominated by noise.

\subsection{Estimation of largest eigenvalues}
The largest eigenvalue $\lambda_\text{max}$ of the Hessian matrix of the PES is an important piece of information for geometry optimizations. It is connected to the initial step size which should be smaller than the reciprocal value of this eigenvalue. $\lambda_\text{max}$ is system specific and typically equals the reciprocal curvature of the PES along bond stretching modes involving the strongest covalent bond.
Based on empirical rules, it is difficult to predict this curvature for complex systems
It however turns out that it is possible to approximate the largest eigenvalue of the hessian matrix using quantities which have to be calculated in the beginning of every geometry optimization. Suppose the geometry optimization is started at $\vb{x}_1$ then $\vb{x}_2 = \vb{x}_1 - \beta \nabla E(\vb{x}_1)$ in most quasi Newton methods. $\lambda_\text{max}$ can be approximated using either of the following two equations:
\begin{gather} \label{eq:approx_ev_power}
    \lambda_\text{max} \approx \frac{ E(\vb{x}_2) - E(\vb{x}_1) + \beta \norm{\nabla E(\vb{x}_1)}^2 }{\frac{1}{2} \beta^2 \norm{\nabla E(\vb{x}_1)}^2} \\
    \lambda_\text{max} \approx \frac{ \norm{ \nabla E(\vb{x}_2) -\nabla E(\vb{x}_1) }}{\beta \norm{\nabla E(\vb{x}_1)}} \label{eq:approx_ev_norm}
\end{gather}
$\lambda_\text{max}$ is an upper bound of both approximations which is presented together with both derivations in \ref{sec:deriv_largest_ev}. Therefore, the larger value of the two expressions is closer to $\lambda_\text{max}$ and its inverse should be used to approximate the initial step size. One still has to choose a value for the parameter $\beta$.
On the one hand side $\beta$ has to be chosen large enough such that the energies and gradients at ${\bf x}_1$ and 
${\bf x}_2$ differ by significantly more than the noise level. On the other hand it has to be chosen smaller than 
the curvature arising from the stretching of the strongest 
bonds found in condensed matter such as the \ce{N2} triple bond.
 A good choice for $\beta$ is $0.1 \unitfrac{Hartree}{Bohr^2}$, respectively $0.001 \unitfrac{eV}{A^2}$.

During the derivation of \cref{eq:approx_ev_power} it is assumed that $x_1$ is in a region of the PES that can be described by a quadratic model centered around a local minimum. In \cref{eq:approx_ev_norm} this assumption is not needed and it should therefore produce more accurate results when the distance between $x_1$ and the nearest local minimum is large.

In our tests \cref{eq:approx_ev_power,eq:approx_ev_norm} the relative error between the approximation and $\lambda_\text{max}$ was usually smaller than 5\%.

\section{Benchmarks}\label{sec:benchmarks}
In order to test the efficiency of the geometry optimizing algorithm presented in \cref{sec:optalg} the performance was compared with the geometry optimizer of \qe{}\cite{qe-2009,qe-2017,qe-2020} and VASP~\cite{VASP_liquid_metal,VASP_amorphous_liquid_metal,VASP_ab_initio_efficiency,VASP_iterative_schemes,VASP_PAW_1,VASP_PAW_2,VASP_LDA}.

\subsection{Benchmark systems}
\begin{figure}[t]
    \centering
    \includegraphics[width=.8\linewidth]{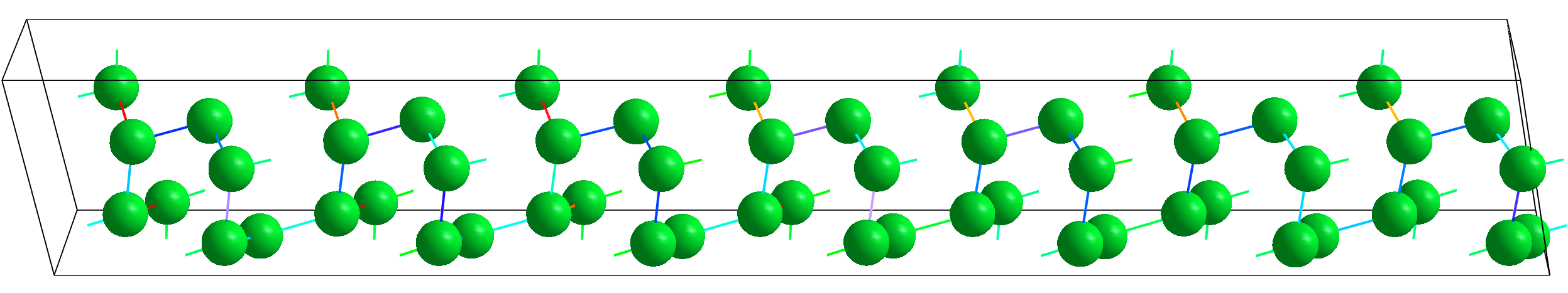}
    \caption{Without our method crystalline cells, whose lattice vectors have very different length, are difficult to optimize.}
    \label{fig:si-long}
\end{figure}

Seven different systems were analyzed in this benchmark.
\begin{itemize}
    \item Carbon in the diamond structure, 8 atoms in lattice cell
    \item Carbon in the diamond structure, 24 atoms in lattice cell
    \item Silicon carbide cell, 8 atoms in lattice cell
    \item Silicon carbide cell, 32 atoms in lattice cell
    \item Long silicon lattice cell containing 56 atoms which is pictured in \cref{fig:si-long}
    \item Formaldehyde, eight formula units
    \item Methylammonium lead iodide (\ce{CH3NH3PbI}) which is described as the delta phase by Flores-Livas et al.~\cite{perovskite-Goedecker} and is also mentioned by Thind et al.~\cite{perovskite-hexagonal-phase}
\end{itemize}
For all of these systems at least 50 geometries were generated using variable cell shape molecular dynamics~\cite{parrinello-md} starting from their respective global minimum. The initial kinetic energy was chosen such that all the structures fall back into the global minimum. This worked in all the benchmark systems except the formaldehyde where some exceptions occurred. These structures were then optimized with our optimizer, the one from \qe{} and the one from VASP.
\begin{table*}
    \begin{minipage}[t]{246pt}
    \begin{tabular}{c|c|cc|cc}
        & $N_{\text{structs}}$ & \multicolumn{2}{c|}{Avg. iterations} & \multicolumn{2}{c}{Avg. length of traject.}   \\ \cline{2-6} 
        &                      & \multicolumn{1}{c|}{vc-SQNM}   & QE    & \multicolumn{1}{c|}{vc-SQNM}       & QE         \\ \hline
\ce{C8}      & 92                   & \multicolumn{1}{c|}{11.7}    & 14.7  & \multicolumn{1}{c|}{1.8}          & 3.3    \\
\ce{C24}     & 50                   & \multicolumn{1}{c|}{12.8}    & 16.0  & \multicolumn{1}{c|}{2.5}          & 4.4    \\
\ce{C8Si8}   & 50                   & \multicolumn{1}{c|}{13.7}    & 16.4  & \multicolumn{1}{c|}{0.8}          & 0.8    \\
\ce{C16Si16} & 50                   & \multicolumn{1}{c|}{28.6}    & 35.0  & \multicolumn{1}{c|}{3.0}          & 4.0    \\
long \ce{Si56} cell & 50            & \multicolumn{1}{c|}{30.7}    & 53.8  & \multicolumn{1}{c|}{42.4}          & 42.6  \\
\ce{C8H16O8} & 50                   & \multicolumn{1}{c|}{61.4}   & 139.8 & \multicolumn{1}{c|}{19.3}          & 43.75 \\
\ce{CH3NH3PbI}   & 50                   & \multicolumn{1}{c|}{141.3}   & 212.5 & \multicolumn{1}{c|}{9.0}         & 13.1   \\
\end{tabular}
    \caption{Results of the benchmark between the optimizer from \qe{} and the vc-SQNM method using energy and forces calculated by \qe{}. $N_{\text{structs}}$ is the number of different structures that were optimized for each system, the average length of the optimization trajectory is given in Bohr.}
    \label{tab:results-benchmarks-qe}
    \end{minipage}\hfill
    \begin{minipage}[t]{246pt}
    \begin{tabular}{c|c|cc|cc}
        & $N_{\text{structs}}$ & \multicolumn{2}{c|}{Avg. iterations} & \multicolumn{2}{c}{Avg. length of traject.} \\ \cline{2-6} 
        &                      & \multicolumn{1}{c|}{vc-SQNM}   & VASP    & \multicolumn{1}{c|}{vc-SQNM}       & VASP       \\ \hline
\ce{C8}      & 92                   & \multicolumn{1}{c|}{12.2}    & 10.0  & \multicolumn{1}{c|}{1.5}          & 2.7      \\
\ce{C24}     & 50                   & \multicolumn{1}{c|}{12.6}    & 10.2  & \multicolumn{1}{c|}{2.1}          & 4.0      \\
\ce{C8Si8}   & 50                   & \multicolumn{1}{c|}{14.2}    & 19.0  & \multicolumn{1}{c|}{1.2}          & 3.1      \\
\ce{C16Si16} & 50                   & \multicolumn{1}{c|}{22.1}    & 36.7  & \multicolumn{1}{c|}{4.1}          & 8.9      \\
long \ce{Si56} cell & 50            & \multicolumn{1}{c|}{36.4}    & 53.7  & \multicolumn{1}{c|}{42.4}       & 93.4    \\
\ce{C8H16O8} & 50                   & \multicolumn{1}{c|}{95.7}   & 580.2 & \multicolumn{1}{c|}{13.05}         & 227.6 \\
\ce{CH3NH3PbI}   & 50                   & \multicolumn{1}{c|}{184.6}   & 482.4 & \multicolumn{1}{c|}{15.4}         & 16.2   \\
\end{tabular}
    \caption{Results of the benchmark between the optimizer from VASP and the vc-SQNM method using energies and forces calculated by VASP. $N_{\text{structs}}$ is the number of different structures that were optimized for each system, the average length of the optimization trajectory is given in Bohr.}
    \label{tab:results-benchmarks-vasp}
    \end{minipage}
\end{table*}

\subsection{Benchmarking method}
In order to make the comparison more meaningful, we used the PES calculated by \qe{} for the vc-SQNM method when it is compared against the geometry optimization algorithm from \qe{}. The same was done for the VASP comparison. An extraordinary high plane wave cutoff of 2200 eV was chosen in order to eliminate inaccuracies that might arise from the change of the plane wave basis which is caused by the varying lattice vectors. When the lattice vectors are varied, either the number of plane waves or the plane wave cutoff must change. For the \qe{} optimizer the default trust radius method was used. In the setup of VASP, the RMM-DIIS optimization algorithm was used. We chose the RMM-DIIS optimization method in VASP since it is stated in the VASP documentation that it is very efficient for structures close to a local minima which is the case in our benchmark\cite{vasp-wiki-geopt}. The step size variable POTIM was set manually to ensure the optimal performance of the algorithm.
To compare the performance of the algorithms two quantities were used. The total number of energy evaluations and the Euclidean length of the optimization trajectory. The latter quantity is of importance for the performance of electronic structure calculations because the self consistent wave function obtained in the previous iteration is typically reused to start the new SCF cycle. Therefore, a short optimization trajectory reduces the total number of SCF cycles.

The periodic geometry optimization method implemented in VASP does not check for convergence of the lattice vectors\cite{vasp-geopt}. Their recommended procedure is to conduct a geometry and lattice optimization simulation, check the convergence of the pressure manually and restart the simulation until the pressure converges. The number of iterations that are obtained in this way for the benchmark are somewhat ambiguous. Because of that we chose a different way to calculate the number of iterations used by VASP. A geometry optimization was stopped when a certain threshold of the norm of the forces was reached. However, this does not guarantee that the pressure is converged in the VASP simulation. Therefore the values in the VASP benchmark are only a lower limit on the number of iterations. In practice, the pressure did however converge in most of the simulations that were done during the benchmarks.

\subsection{Performance of our method with noisy forces}
Since this method may be used in situations where calculating accurate derivatives may be difficult and forces therefore contain a lot of noise, we want to demonstrate how geometry optimizations are affected by noisy forces. This is done by randomly displacing atoms from a the ground configuration of bulk silicon with 64 atoms in the unit cell. Energies, forces and lattice derivatives are calculated using the force field from Bazant et al.~\cite{silicon-ff}. In this force field, the analytic expression of the forces is known. Therefore, the forces and lattice derivatives can be calculated exactly. To simulate an environment where the derivatives contain noise a normal distributed random number with mean zero and a standard deviation of $10^{-4} \unitfrac{Hartree}{Bohr}$ is added to each force and lattice derivative component. In order to see the effect of the artificial noise on the efficiency of the geometry optimization algorithm, the convergence of the maximal force and lattice derivative component is compared against a geometry optimization that starts at the same initial configuration but the noise is not added to the forces.

\begin{figure}
    \centering
    \includegraphics{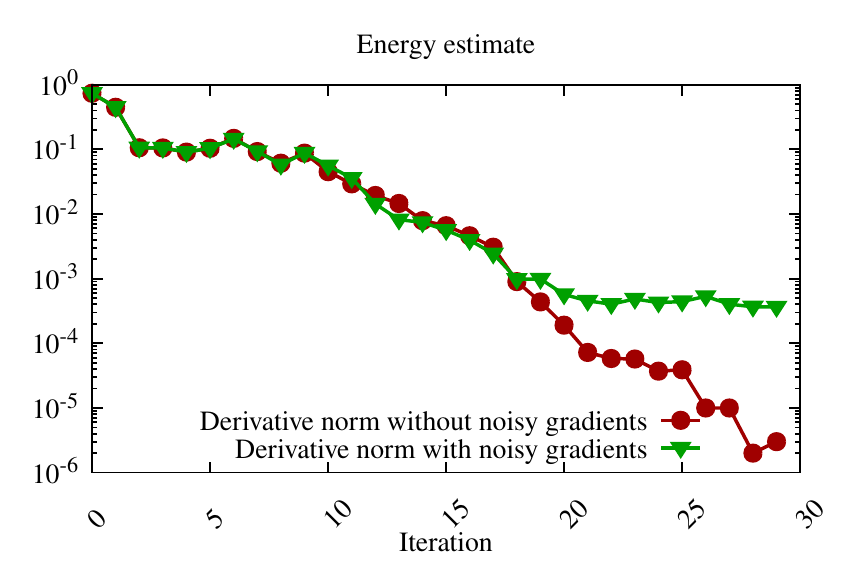}
    \caption{The maximal absolute value of forces and lattice derivatives with noise added to them is shown in green color and without noise in red color.}
    \label{fig:noisy-forces}
\end{figure}

\cref{fig:noisy-forces} shows the convergence of the force norm with and without noise. When the convergence of the geometry optimization with noisy forces is compared to the same geometry optimization without noisy forces, two things can be seen. First, the force norms can not be brought below the limit of 3 to 5 times  $\sigma$. Secondly, the convergence rate of the geometry optimization starts to slow down only when this limit is reached. This experiment was repeated numerous times and the noisy forces always impacted the geometry optimizations as described above. This observation is especially useful in combination with the result of \cref{sec:noise-force} where we provide a way to estimate the standard deviation of the noise term present in the forces which can be used as a stopping criterion for the geometry optimization.

\subsection{Results}
The results of the performance comparison are displayed in \cref{tab:results-benchmarks-qe,tab:results-benchmarks-vasp}. In most of the comparisons, the vc-SQNM method clearly outperforms the methods implemented in VASP and \qe{}. The average length of the optimization trajectory is shorter in every system that is considered here which can speed up DFT calculations.

The long silicon cell and the formaldehyde system are the most difficult systems to optimize included in the benchmark. In these difficult systems, the difference in performance between the vc-SQNM method and the VASP and \qe{} implementations is particularly larger. As expected, the vc-SQNM method, that uses the transformed coordinates $\vb q_i$ together with $\tilde{A}$, can handle difficult systems better than competing algorithms.

\section{Conclusion}
We introduced a transformation that improves the condition number for variable cell shape optimizations.
The vc-SQNM method is compared against the optimization methods from VASP and \qe{} in an extensive benchmark. It is shown that it can outperform the optimizer from \qe{} and VASP in almost all cases that were investigated. 
Especially for difficult system the number of required energy and force evaluations can be decreased. In all of the systems tested in the benchmark, the average euclidean length of the optimization trajectory was significantly shorter which will reduce the number of steps in the SCF cycle and therefore the computational effort.
The high stability of our method is witnessed by the fact that 
over 300'000 periodic geometry optimization were successfully completed in studies within group\cite{minhop_CHS,minhop_high_sym} where the minima hopping method is used\cite{mh,Amsler2010,Sicher2011,Roy2008}.

Upon publication of this manuscript we will publish an implementation of our method on GitHub: \url{https://github.com/}.

\section*{Acknowledgments} 
The calculations were performed on the computational resources of the Swiss National Supercomputer (CSCS) under project s1167 and at sciCORE (\url{http://scicore.unibas.ch/}) scientific computing center at University of Basel. Financial support was obtained from the Swiss National Science Foundation.

\appendix
\section{Derivation of the proper side length of the supercell}\label{sec:deriv_proper_sidlength}
The modified lattice coordinates have the form $ a = \frac{\tilde{a} \cdot a_0}{k} $, $a_0$ is the side length of the initial cube. The energy of the modified PES is given by 
$
    \tilde{E} = \tilde{E}(\vb r_1, \dots, \vb r_{\nat}, \frac{\tilde{a} \cdot a_0}{k})
$.
The first and second derivative are given by:
\begin{equation}\label{eq:mod_sec_deriv_append}
    \pderiv{\tilde{E}}{\tilde{a}} = \pderiv{a}{\tilde{a}} \pderiv{}{a} E \left( \vb r_1, \dots, \vb r_{\nat}, \frac{\tilde{a} a_0}{k} \right) = \frac{a_0}{k} \pderiv{E}{a}, \quad
    \psderiv{\tilde{E}}{\tilde{a}} = \left(\pderiv{a}{\tilde{a}} \right)^2 \psderiv{E}{a} = \frac{a_0^2}{k^2} \underbrace{\psderiv{E}{a}}_{\sim \nat^{\frac{1}{3}}} \sim \frac{a_0^2}{k^2} \nat^{\frac{1}{3}}
\end{equation}
A good approximation for $a_0$ is $n \cdot l = \nat^{\frac{1}{3}} \cdot l $ where $l$ is the equilibrium bond length and $n$ is the number of atoms per side length of the cube. Using the fact that the eigenvalues of cubic cells are proportional to $\nat^{\frac{1}{3}}$ and setting the second derivative constant one gets:
\begin{equation}
    \psderiv{\tilde{E}}{\tilde{a}} \sim \frac{\left(\nat^{\frac{1}{3}}\right)^2}{k^2} \nat^{\frac{1}{3}} = \frac{\nat}{k^2} = const. \quad \implies k \sim \sqrt{\nat}.
\end{equation}
\section{Derivation largest eigenvalue approximations}\label{sec:deriv_largest_ev}
Derivation of $\lambda_\text{max} \approx \frac{ E(\vb{x}_2) - E(\vb{x}_1) + \beta \norm{\nabla E(\vb{x}_1)}^2 }{\frac{1}{2} \beta^2 \norm{\nabla E(\vb{x}_1)}^2}$:\\
The power iteration can be used to approximate the largest eigenvalue of a matrix $A$ and an arbitrary non-zero vector $\vb{b}$: $\lambda_{\text{max}} = \lim_{k \rightarrow \infty} \frac{\vb{b}^T \cdot A^{2k + 1} \cdot \vb{b}}{\norm{A^k \vb{b}}^2} \implies \lambda_{\text{max}} \approx \frac{\vb{b}^T A^3 \vb{b}}{\norm{A \vb{b}}^2}$. Let $E(\vb{x})$ be the taylor expansion of the PES around a local minimum with coordinates $\vb{x}_0$ up to second order: 
\begin{align}
    E(\vx) &= E(\vx_0) + \underbrace{(\vx - \vx_0)^T \nabla E(\vx_0)}_{=0} + \frac{1}{2} (\vx - \vx_0) ^T H (\vx - \vx_0) \\
    \nabla E (\vx) &= H (\vx - \vx_0).
\end{align}
Consider $E(\vx - \beta \nabla E(\vx)) = E(\vx) -\beta \nabla E(\vx)^T \nabla E(\vx) + \frac{1}{2} \beta^2 \nabla E(\vx)^T H \nabla E(\vx)$. Using the power iteration approximation 
\begin{align}
    \lambda_\text{max} &\approx \frac{(\vx - \vx_0)^T H^3 (\vx-\vx_0)}{\norm{H(\vx-\vx_0)}^2} = \underbrace{\frac{\nabla E(\vx)^T H \nabla E(\vx)}{\norm{\nabla E(\vx)}^2}}_{\leq \lambda_\text{max}} \nonumber \\
    &= \frac{ E(\vx - \beta \nabla E(\vx)) - E(\vx) + \beta \norm{\nabla E(\vx)}^2 }{\frac{1}{2}\beta^2 \norm{\nabla E(\vx)}^2}. \label{eq:power_it_deriv}
\end{align}
Derivation of $\lambda_\text{max} \approx \frac{ \norm{ \nabla E(\vb{x}_2) -\nabla E(\vb{x}_1) }}{\beta \norm{\nabla E(\vb{x}_1)}}$:\\
Let $E(\vb{x})$ be the taylor expansion of the PES around an arbitrary point $\vb{x}_0$ up to second order: 
\begin{align}
    E(\vx) &= E(\vx_0) + (\vx - \vx_0)^T \nabla E(\vx_0) + \frac{1}{2} (\vx - \vx_0) ^T H (\vx - \vx_0) \\
    \nabla E (\vx) &=\nabla E (\vx_0) +  H (\vx - \vx_0).
\end{align}
Consider $\nabla E(\vx - \beta \nabla E(\vx)) = \nabla E (\vx_0) + H (x - \beta \nabla E(x)- x_0) = \nabla E(x) - \beta H \nabla E(x) $. Therefore, $\norm{\nabla E(\vx - \beta \nabla E(\vx)) - \nabla E(\vx)} = \beta \norm{H \nabla E(\vx)} \leq \beta \norm{H} \norm{\nabla E(\vx)} \leq \beta \lambda_\text{max} \norm{\nabla E(\vx)}$ because the Frobenius norm is consistent with the Euclidean vector norm. Rearranging the terms results in the following approximation: 
\begin{equation}
    \lambda_\text{max} \approx \frac{ \norm{ \nabla E(\vx - \beta \nabla E(\vx)) -\nabla E(\vx) }}{\beta \norm{\nabla E(\vx)}}.
\end{equation}

\bibliographystyle{elsarticle-num} 
\bibliography{main}

\end{document}